\documentclass[aps,prl,twocolumn,groupedaddress,floatfix]{revtex4}

\usepackage{graphicx}
\usepackage{dcolumn} 
\usepackage{bm}      

\begin{document}

\title{Observation of pseudogap behavior in a strongly interacting Fermi gas}
\author{J. P. Gaebler$^{1}$, J. T. Stewart$^1$, T. E. Drake$^1$ and D. S. Jin$^1$, A. Perali$^2$, P. Pieri$^2$ and G. C. Strinati$^2$}

\affiliation{$^1$ JILA, National Institute of Standards and Technology
(NIST) and University of Colorado, Department of Physics, University
of Colorado, Boulder, CO 80309, USA}

\affiliation{$^2$ Dipartimento di Fisica, Universit$\grave{a}$ di Camerino, I-62032 Camerino, Italy}

\date{\today}

\pacs{??}

\maketitle


\textbf{Ultracold atomic Fermi gases present an opportunity to study
strongly interacting Fermi systems in a controlled and uncomplicated
setting. The ability to tune attractive interactions has led to the
discovery of superfluidity in these systems with an extremely high
transition temperature \cite{Regal2006, Ketterle2008}, near
$\frac{T}{T_F} = 0.2$. This superfluidity is the electrically
neutral analog of superconductivity; however, superfluidity in
atomic Fermi gases occurs in the limit of strong interactions and
defies a conventional BCS description. For these strong
interactions, it is predicted that the onset of pairing and
superfluidity can occur at different temperatures \cite{Randera1998,
Perali2002, Chen2006}.  This gives rise to a pseudogap region where,
for a range of temperatures, the system retains some of the
characteristics of the superfluid phase, such as a BCS-like
dispersion and a partially gapped density of states, but does not
exhibit superfluidity.  By making two independent measurements: the
direct observation of pair condensation in momentum space and a
measurement of the single-particle spectral function using an analog
to photoemission spectroscopy \cite{Stewart2008}, we directly probe
the pseudogap phase.   Our measurements reveal a BCS-like dispersion
with back-bending near the Fermi wave vector $k_F$ that persists
well above the transition temperature for pair condensation.}

In conventional superconductors, fermion pairs and superconductivity
appear simultaneously at $T_c$. The single-particle, or fermionic,
excitation spectrum of a conventional superconductor follows a BCS
dispersion given by
\begin{equation}\label{eq:BCSdispersion}
E_s=\mu \pm \sqrt{(\epsilon_k-\mu)^2+\Delta^2},
\end{equation}
where $\epsilon_k = \hbar^2 k^2/2m$, $\mu$ is the chemical potential,
and $\Delta$ is the superfluid order parameter. The lower branch of
the dispersion (minus sign in Eqn. 1), which is the occupied one at
low temperature, has a positive slope at low momentum and then turns
around and has a negative slope at high momentum. This
``back-bending" behavior arises because of the excitation gap and is
a characteristic signature of superconductivity. In unconventional
superconductors, such as high Tc superconductors, this back-bending
in the dispersion has been observed both below \cite{Damascelli2004}
and, remarkably, above $T_c$ \cite{Kanigel2008}. The observation of
back-bending above $T_c$ represents a dramatic departure from
conventional BCS theory, which predicts that the normal state above
$T_c$ is a Fermi liquid with a monotonically increasing
single-branch dispersion and a density of states that is smooth
through the Fermi surface. The departure from a conventional BCS
description above $T_c$ in the form of a gapped excitation spectra
is the essence of the pseudogap.

\begin{figure*}
\includegraphics[width=\linewidth]{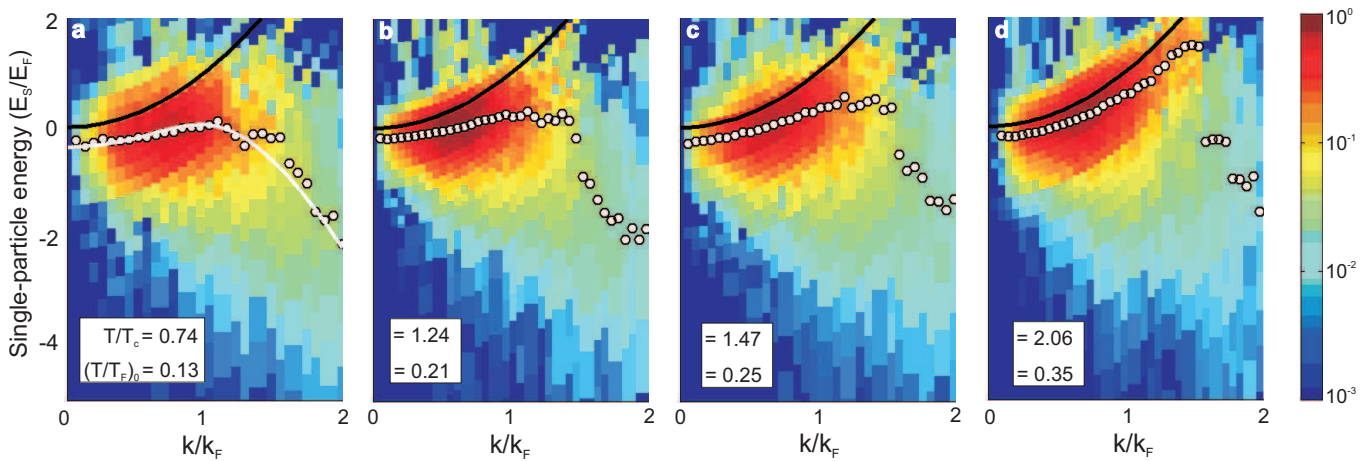}
\caption{\textbf{Photoemission spectra throughout the pseudogap
regime.} Spectra are shown for Fermi gases at four different
temperatures, each with an interaction strength characterized by
$(k_Fa)^{-1} \approx 0.15$. The intensity plots show the fraction of
out-coupled atoms as a function of their single-particle energy
(normalized to $E_F)$ and momentum (normalized to $k_F$), where
$E=0$ corresponds to a non-interacting particle at rest.  The
spectra are normalized so that integrating them over momentum and
energy gives unity. White dots indicate the centers extracted from
gaussian fits to individual energy distribution curves (traces
through the data at fixed momentum).  The black curve is the
quadratic dispersion expected for a free particle.  \textbf{a} At $T=
0.74\hspace{.7mm}T_c$, we observe a BCS-like dispersion with
back-bending, consistent with previous measurements
\cite{Stewart2008}. The white curve is a fit to a BCS-like
dispersion, Eqn. 1. \textbf{b,c} At $T= 1.24
\hspace{.7mm}T_c$ and $T= 1.47 \hspace{.7mm}T_c$, respectively, the
dispersion with back-bending persists even though there is no longer
any superfluidity.   \textbf{d} At $T= 2.06 \hspace{.7mm}T_c$, the
dispersion does not display back-bending in the range of $0 < k <
1.5 \hspace{.7mm} k_F$.  In all the plots there is a
negative dispersion for $k/k_F > 1.5$.  We attribute this weak feature (note the
log scale) to a $1/k^4$ tail and not to the gap.
}\label{PESdata}
\end{figure*}

A satisfactory explanation of the origin and nature of the observed
pseudogap phase in high Tc superconductors has remained elusive due
to the complexity of the materials. In contrast, ultracold atomic
gases are relatively free of complexity, for example, having no
underlying lattice structure, impurities, or domain boundaries.
Moreover, the interactions responsible for pairing and superfluidity
in ultracold atom gases are well understood at the few-body level.
As a result, these systems are ideally suited for investigating the
prediction of a pseudogap phase due to pre-formed pairs. There is
much scientific literature on the topic of the pseudogap in strongly
interacting atom gases with wide-ranging viewpoints and conclusions
\cite{Janko1997, Yanase1999, Perali2002, Bruun2006, Massignan2008,
Barnea2008, Magierski2009, Chen2009b, Tsuchiya2009, Randeria92,
Melo93}, including a recent article that predicts no pseudogap phase
at all \cite{Haussmann2009}.  In some theories, the pseudogap phase
is predicted to have a BCS-like dispersion (Eqn. 1), but where
$\Delta$ is no longer the superfluid order parameter but instead
corresponds to an excitation gap due to the formation of incoherent
pairs \cite{Janko1997, Yanase1999, Perali2002, Bruun2006,
Massignan2008, Barnea2008, Magierski2009, Chen2009b, Tsuchiya2009,
Randeria92, Melo93}.  However, for the atomic gases, there is not
yet experimental data to establish the existence of a pseudogap
phase and confirm its properties. Radio frequency (rf) spectroscopy
experiments that probe excitations have been performed above and
below the critical temperature \cite{Chin2004a,Schirotzek2008b}, but
their interpretation relies on assuming a specific dispersion
relation and therefore they cannot be used to distinguish between a
pseudogap and a normal phase \cite{Massignan2008, Haussmann2009}.

The question is then, do we have the necessary measurement tools to
look for pseudogap physics in the neutral atom gas system? To probe
the defining properties of a pseudogap regime one needs both a
measurement of the transition temperature as well as a probe of the
single-particle excitation spectra. In the atomic gas system, the
onset of the superfluid phase is clearly detected through the
observation of momentum-space condensation of atom pairs
\cite{Regal2004a}. To probe the single-particle excitation spectra,
we use a technique recently developed for atoms that uses
momentum-resolved rf spectroscopy to realize an analog of
photoemission spectroscopy \cite{Stewart2008}. Using these two
measurements: the direct observation of pair condensation to
determine $T_c$, and, momentum-resolved rf spectroscopy to probe the
pairing gap, we can now explore the issue of the pseudogap in atomic
systems. In this paper, we report that a BCS-like dispersion, with
back-bending near $k_F$, indeed persists even for temperatures
substantially above the measured critical temperature for
superfluidity. For the atomic gas system, which is clean and simple
in comparison to high Tc materials, this result demonstrates the
existence of a pseudogap region where incoherent pairs of correlated
fermions exist above $T_c$.

To perform these experiments, we cool a gas of fermionic $^{40}$K
atoms to quantum degeneracy in a far detuned optical dipole trap as
described in previous work \cite{Stewart2008}. We obtain a $50/50$
mixture of atoms in two spin states, namely the $|f,m_f\rangle=
|9/2, -9/2\rangle$ and $|9/2,-7/2\rangle$ states, where $f$ is the
total atomic spin and $m_f$ is the projection along the
magnetic-field axis. Our final stage of evaporation occurs at a
magnetic field of 203.5 G, where the $s$-wave scattering length that
characterizes the interactions between atoms in the $|9/2,
-9/2\rangle$ and $|9/2,-7/2\rangle$ states is approximately 800
$a_0$, where $a_0$ is the Bohr radius.  At the end of the
evaporation we increase the interactions adiabatically with a slow
magnetic-field ramp to a Feshbach scattering resonance.

To vary the temperature of the atom cloud, we either truncate the
evaporation or parametrically heat the cloud by modulating the
optical dipole trap strength at twice the trapping frequency. To
determine the temperature of the Fermi gas we expand the weakly
interacting gas and fit the momentum distribution to the expected 2D
distribution \cite{Regal2006} and extract $(T/T_F)_0$, where the
subscript $(_0)$ indicates a measurement made in the weakly
interacting regime, before ramping the magnetic field to the
Feshbach resonance. For the data presented here, we obtain clouds at
final temperatures ranging from $(T/T_F)_0 = 0.12$ to $0.43$ with
$N= 1\times 10^5$ to $1.8 \times 10^5$ atoms per spin state. The
trap frequencies vary depending on the final intensity of the
optical trap and range from $180$ to $320$ Hz in the radial
direction and $18$ to $27$ Hz in the axial direction.
Correspondingly, the Fermi energy, $E_F$, ranges from $h \cdot 8$
kHz to $h \cdot 13$ kHz, where $h$ is Planck's constant. The Fermi
energy is obtained from $N$ and the geometric mean trap frequency,
$\nu$, as $E_F = h \nu (6N)^{1/3}$.  We define the
Fermi wave vector as $k_F=\frac{\sqrt{2mE_F}}{\hbar}$ and the Fermi
temperature as $T_F=E_F/k_B$, where $k_B$ is the Boltzmann constant.
It is important to note that the trapped gas has a spatially
inhomogeneous density, and one can define a local Fermi energy, and
corresponding local Fermi wave vector, that vary across the cloud.

\begin{figure}
\includegraphics[width=\linewidth]{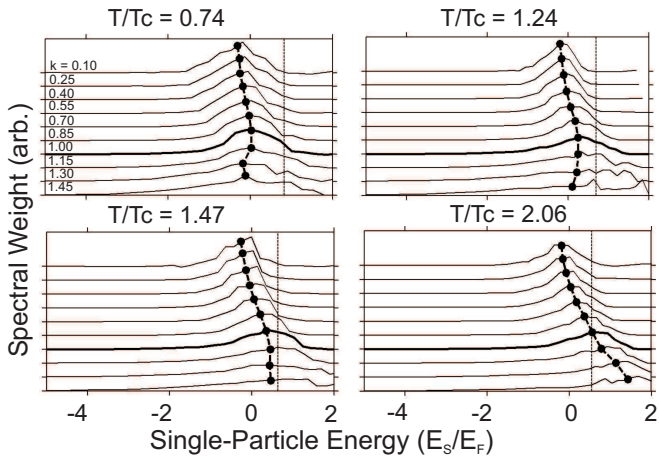}
\caption{\textbf{Energy Distribution Curves (EDCs)} EDCs are
obtained by taking vertical traces at fixed $k$ through the
photoemission spectra shown in Fig. 1. We show EDCs between
$\frac{k}{k_F} = 0.1$ (top) and $\frac{k}{k_F} = 1.4$ (bottom) for
the four data sets with $T/T_c$ labeled above each figure. Each
plotted EDC is an average of EDCs over a range of approximately
$0.15 k_F$. The EDC at $\frac{k}{k_F}= 1.0$ is shown in bold. Black
dots indicate the centers of the gaussian fits to the EDCs.  Each
EDC is normalized to have an area of unity.  Vertical dotted lines
are placed at the local $E_F$ that corresponds to the estimated
average density of the gas.} \label{AllEDCs}
\end{figure}

Momentum-resolved rf spectroscopy realizes photoemission
spectroscopy for strongly interacting atoms \cite{Stewart2008}, much
like angle-resolved photoemission spectroscopy (ARPES) for strongly
correlated electron systems. In this spectroscopy, an rf photon
flips the spin of an atom to a third hyperfine spin state and then
the spin-flipped atoms are counted as function of their momentum. As
in ARPES \cite{Damascelli2004}, conservation of energy and momentum
are used to extract the energy and momentum of the fermion (which in
our case is an entire atom) in the strongly correlated system. A key
feature of this measurement is that the spin-flipped atoms are
``ejected" from the system in the sense that they have only very
weak interactions with the other atoms. This means that the
spin-flipped atoms have the usual free-particle dispersion, and
moreover, their momentum distribution can be measured using
time-of-flight absorption imaging with no significant effects of
interactions or collisions on the ballistic expansion. This
technique was recently applied to a gas just below $T_c$ and
revealed a BCS-like back-bending dispersion characteristic of an
excitation gap \cite{Stewart2008}.

To perform the photoemission experiments on atoms, we turn on a
short rf pulse to transfer atoms from the $|9/2,-7/2\rangle$ state
to the unoccupied and weakly-interacting $|9/2,-5/2\rangle$ state.
We then immediately turn off the trap and state-selectively image
the out-coupled atoms on a CCD camera after time-of-flight expansion. The
rf pulse is kept much shorter than a trap period to ensure that the
momentum of the out-coupled atoms does not change. The length of the rf pulse limits our energy resolution to approximately $0.2 E_F$. As described in
our previous work, the intensity of atoms out-coupled as a function
of momentum for each rf frequency can be used to reconstruct the
occupied single-particle states \cite{Stewart2008}. With this
information, one can determine the occupied part of the Fermi
spectral function and probe the energy dispersion.  It is important to note that
 unlike ARPES experiments in condensed matter physics the value of the chemical potential
 is not determined in this experiment.  Rather, in our plots zero energy corresponds to the
 energy of a non-interacting atom at rest.

We present our photoemission spectroscopy data studying the
pseudogap of a strongly interacting Fermi gas in figures 1 and 2.
The dimensionless parameter that characterizes the interaction
strength for this data is $1/k_F a = 0.15(3)$, where $a$ is the
$s$-wave scattering length.  In Fig. 1, we plot the fraction of
out-coupled atoms as a function of their single-particle energy and
momentum for temperatures encompassing the pseudogap regime. In the
intensity plots, white dots indicate the centers derived from
unweighted gaussian fits to each of the energy distribution curves,
or EDCs, (vertical trace at a given wave vector). The energy
dispersion mapped out with these fits (white dots) can be contrasted
to the expected free particle dispersion for an ideal Fermi gas
(black curve).  In Fig. 2 we show the same data plotted as EDCs for
wavevectors ranging from $k/k_F = 0.1$ to $k/k_F = 1.4$. In order to
show the evolution of the spectral function from below $T_c$ through
the pseudogap regime the data are shown for four temperatures,
$(T/T_F)_0 = 0.13, 0.21, 0.25$ and $0.35$.

For the data below $T_c$ (Fig. 1a), we see a smooth back-bending
that occurs near $k=k_F$.  The white curve in Fig. 1a shows a
BCS-like dispersion curve, Eqn. 1, discussed above; here, we fit to
the white dots for momenta in the range $0 < k < 1.4 \hspace{.7mm}
k_F$. While we cannot use this fit to extract the gap and chemical
potential in a model-independent way due to the harmonic trapping
confinement, the BCS-like fit is consistent with a large pairing
gap, on order of $E_F$, as expected for a Fermi gas near the center
of the BCS-BEC crossover \cite{Regal2006, Ketterle2008}.

A striking feature of the data is that the measured spectral
function evolves smoothly as the temperature is increased above
$T_c$.  In fact, in all four of our data sets in Fig. 1, we observe
a weak signal with a strong negative dispersion at high momenta.  It
has been recently pointed out that one expects universal behavior at
$k>>k_F$ for a Fermi gas with short-range, or contact, interactions
\cite{Tan08}, and, moreover, that this will give rise to a weak,
negatively dispersing feature in the Fermi spectral function
\cite{Schneider2009b}.  Recently, we have directly verified this
universal behavior with measurements of the momentum distribution
and found empirically that the expected $1/k^4$ tail occurs for
$k>1.5k_F$ \cite{Stewart2010}. Therefore, we attribute the
negative dispersion seen at large $k$ to this universal behavior for
contact interactions.  While the strength of this feature should
reflect the state of the system, the negative dispersion for
$k>1.5k_F$ does not, by itself, provide evidence of a BCS-pairing
gap \cite{Schneider2009b}.

In the case of a pairing gap, we expect the spectral function to
exhibit back-bending for $k$ near $k_F$.  To avoid effects of the
universal behavior at large $k$, we consider the spectral function
for $k<1.5k_F$.  For the three lowest temperatures, we observe a
BCS-like dispersion with back-bending behavior that occurs near
$k/k_F = 1$. We interpret this as evidence for the existence of a
pseudogap regime above $T_c$ comprised of uncondensed pairs in the
strongly interacting Fermi gas.  At our highest temperature,
$(T/T_F)_0 = 0.35$, we observe a dispersion that increases
quadratically through the region of $k/k_F = 1$ and then appears to
jump discontinuously near $k/k_F = 1.5$.  At the same point, the
amplitudes of the Gaussian fits to the EDCs also drop sharply (see
inset to Fig. 3). Thus, we conclude that, as the occupation of the
positively dispersing feature vanishes, the fits jump to a
distinct, lower energy feature in the spectral function.  As
discussed above, this lower energy feature is consistent with the
predicted effect of universal behavior at large $k$ on the spectral
function \cite{Schneider2009b}.

In general, for data taken at finite temperature but still in the
pseudogap regime one might expect to see population in the excited
branch of the BCS Bogoliubov dispersion (plus sign in Eqn. 1).
Signal in this branch represents thermally populated excitations
above the pairing gap. The data in the region of $1 < \frac{k}{k_F}
< 1.5$ are suggestive of some occupation in this branch. However,
the limited signal-to-noise ratio makes it difficult to identify the
excited branch in our data. In addition, the inhomogeneous density
of the trapped gas could make the observation of two distinct
branches more difficult. To be conservative, we fit each of the EDCs
to a single gaussian, and we find this to be sufficient to identify
back-bending. It will be a subject of further research to see if the
excited branch can be more clearly observed in a momentum-resolved
atom photoemission measurement.

\begin{figure}
\includegraphics[width=\linewidth]{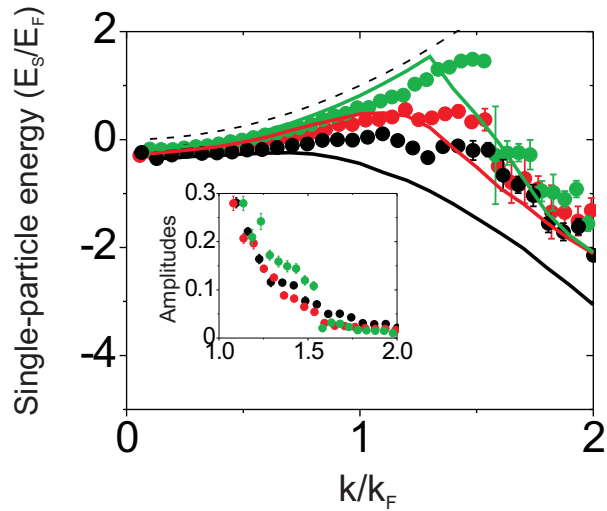}
\caption{\textbf{Single-particle dispersion curves.}The fits to the
EDC centers are shown for the three temperatures in Fig. 1 and Fig.
2 a,c and d, represented by black, red, and green, respectively.  We
observe a BCS-like dispersion, smooth back-bending near
$\frac{k}{k_F}=1$, for temperatures below and moderately above
$T_c$. For the highest temperature, we observe a quadratic dispersion
near $\frac{k}{k_F}=1$ and a sharp discontinuity near
$\frac{k}{k_F}=1.5$.  The lines are theory curves that include
effects of the harmonic trap and contact interaction, as described
in the text. \textbf{Inset} We show the amplitudes from the gaussian
fits to the EDCs for the same experimental data.  The fit amplitudes
evolve smoothly for the lower temperatures but jump discontinuously
for the highest temperature gas.} \label{Dispersions}
\end{figure}

In Fig. 3, we directly contrast the dispersions obtained for
temperatures $T/T_c = 0.74$, $1.47$, and $2.06$, shown in black, red,
and green, respectively. We compare the experimental dispersions
(circles) to a BCS-BEC crossover theory described in Ref.
\cite{Perali2002}. To compare to the experimental data, we fit
theoretical EDCs to single gaussians to extract the centers; the
results are shown as lines in Fig. 3. The theory incorporates the
trapping confinement as well as the energy resolution due to the
finite rf pulse duration. The theory, which gives the expected
$k^{-4}$ behavior at high $k$ for the momentum distribution, agrees
qualitatively with the experimental data. Both experimental and
theoretical dispersions show smooth BCS-like dispersions with
back-bending near $k = k_F$ for temperatures up to $T/T_c = 1.47$.
For $T/T_c = 2.06$, both the experimental and theory show a
quadratic dispersion before the signal decays around $k = 1.5 k_F$,
leaving a much weaker negatively dispersing feature as predicted for
a normal gas with contact interactions. The apparent disagreement between theory and experiment at $T/T_c = 0.74$ can be attributed to a sharp variation of the order parameter with temperature close to $T_c$. In the inset of Fig. 3, we
show the amplitudes of the gaussian fits to the measured EDCs for
each of the three temperatures.

With theoretical calculations for a homogeneous Fermi gas, we find
that the strongly interacting gas with pre-formed pairs and a normal
Fermi liquid have distinct spectral functions.  Namely, the paired
state shows a smooth avoided crossing (such as described by Eqn. 1)
while the normal Fermi liquid exhibits a sharp crossing leading to a
cusp or apparent discontinuity in the occupied part of the spectral
function.  The smooth behavior in the measured dispersion at the
three lower temperatures, and the sharp jump in the dispersion at
large $k$ for the highest temperature data are consistent with this
theoretical picture.

\begin{figure}
\includegraphics[width=\linewidth]{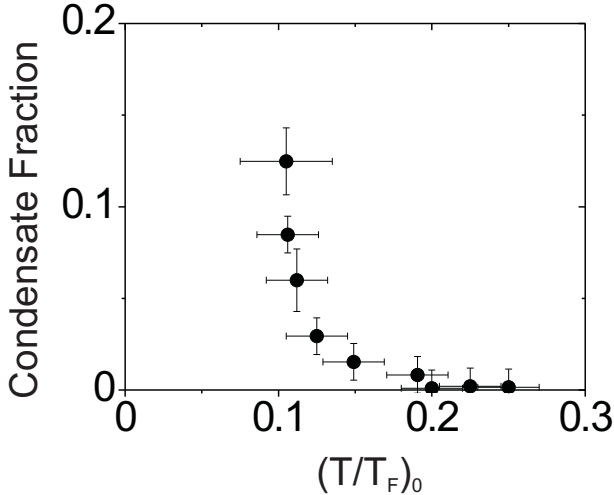}
\caption{\textbf{Condensate fraction as a function of temperature.}
Using time-of-flight expansion at $B = 202.1$ G where our
photoemission experiments are performed, we map out the condensate
fraction. Temperature is measured in the weakly interacting regime
before the adiabatic ramp to strong interactions. We find
$(\frac{T_c}{T_F})_0=0.17\pm0.02$.  Note that the density of the
trapped cloud decreases with increasing distance from the trap
center, and therefore, in a local density picture, even at
$\frac{T}{T_F} = 0.17$ only the part of the gas at the very center
of the trap is below $T_c$.
 } \label{CondFrac}
\end{figure}

To determine $T_c$, we probe pair condensation in our atomic Fermi
gas following the procedure introduced in Ref. \cite{Regal2004a}.
This technique directly probes coherence and has been used to map
out $T_c$ as a function of temperature and interaction strength
\cite{Regal2004a}. The fact that the Bose condensation of fermion
pairs corresponds to a superfluid phase transition was demonstrated
unambiguously with the observation of a vortex lattice in a rotated
Fermi gas below $T_c$ \cite{Zwierlein2005b}. In addition, the
accuracy of the condensate fraction measurements has been
investigated both theoretically \cite{Perali2005, Goral2008} and
experimentally \cite{Zwierlein2005}. In Fig. 4, we show the measured
pair condensate fraction as a function of the initial temperature of
the Fermi gas. As an empirical definition of $T_c$, we use the
temperature where the measured condensate fraction is 1 $\%$.  The
value of 1 $\%$ is chosen because, when testing our fits with
simulated data, we find that we cannot differentiate between a Bose
distribution above $T_c$ and one with a 1 $\%$ condensate fraction.
We find $T_c=(0.17\pm0.02) \hspace{0.7mm} T_F$ at $1/k_F a =
0.15(3)$, and using this we report $T/T_c$ for our photoemission
spectroscopy data.

Contrasting the photoemission spectroscopy data with this direct
measure of the temperature $T_c$ below which the system has coherent
pairs, we find that BCS-like back-bending persists well above $T_c$
in what we identify to be the pseudogap phase. Above the superfluid
transition temperature, these strongly interacting Fermi gases are
clearly not described by a Fermi liquid dispersion and the existence of
many-body pairing well above $T_c$ marks a significant departure
from conventional BCS theory. It is intriguing to note that our
measurements are qualitatively similar to ARPES results in high Tc
superconductors \cite{Kanigel2008}, even though the atomic Fermi gas
is a much simpler system that does not even have an underlying
lattice structure.  However, high Tc materials and ultracold atom systems differ substantially and for example it may  be important to consider that the atomic Fermi gas superfluids have a higher $\frac{T_c}{T_F}$ compared to high Tc superconductors and are more clearly in the region of the BCS-BEC crossover.

\section{Methods}
\textbf{Feshbach resonance location}\\
 To create a strongly interacting gas we ramp the
magnetic field after evaporation to a value of 202.1 G at a inverse
ramp rate of 14 ms/G. The data presented here is at the same
magnetic field as previous ``on resonance" results presented in Fig.
3b of Ref. \cite{Stewart2008}, where the value of $a$ was based on a
measurement of the resonance position in Ref. \cite{Regal2004a}.
However, from a new measurement based on molecule binding energies
determined from rf spectra, we find the resonance position to be
$B_0 = 202.20(2)$ G and width to be $w=7.1(2)$ G. With the new
resonance parameters and $B= 202.1$ G, we find that the
characteristic dimensionless interaction parameter $1/k_F a$ is
$0.15(3)$. This corresponds to the region of the BCS-BEC crossover
where the gas is extremely strongly interacting and the superfluid
gap is expected to be on order of $E_F$. Note that in the absence of
many-body physics, the two-body prediction of the molecule binding
energy at 202.1 G is 480 Hz, which is less than $0.05 E_F$.

\textbf{Photoemission spectroscopy}\\
 For the work presented here, we have
improved the signal-to-noise ratio of the photoemission spectra by a
factor of four compared to our previous measurement
\cite{Stewart2008}. Previously, a limitation to the signal-to-noise
was related to imaging the out-coupled $|9/2,-5/2\rangle$ atoms,
which lack a closed cycling transition for absorption imaging. Now,
we transfer the out-coupled atoms to the $|9/2,-9/2\rangle$ state
with two rf $\pi$-pulses. Because the number of atoms in the
$|9/2,-5/2\rangle$ state is relatively small, this requires that we
first optically pump the atoms remaining in the $|9/2,-7/2\rangle$
and $|9/2,-9/2\rangle$ states to another hyperfine manifold. In this
way, we can image the out-coupled atoms with the cycling transition
for the $|9/2,-9/2\rangle$ state without contamination from the much
larger population of atoms that were unaffected by the rf
spectroscopy. Before constructing the photoemission spectra, we
clean up the raw images by setting to zero data at large radii where
the signal drops below technical noise.

\textbf{Density inhomogeneity of the trapped gas}\\
One can define a local Fermi energy, and corresponding local Fermi
wave vector, that vary across the cloud. We can estimate average
density of the strongly interacting gas by taking the average
density of an ideal trapped Fermi gas at a particular
$\frac{T}{T_F}$ and multiplying by
$(\frac{E_{pot}}{E^0_{pot}})^{-3/2}$.   Here,
$\frac{E_{pot}}{E^0_{pot}}$ is the measured ratio (at finite $T$) of
the potential energy of the strongly interacting gas to that of a
non-interacting gas \cite{Stewart2006}.  For the the data shown in
Fig. 1 a-d, the local Fermi energy, in units of the previously
defined $E_F$, that corresponds to this average density is
$0.81,0.69,0.62,$ and $0.53$, respectively.  The corresponding local
Fermi wave vector, in units of $k_F$, is $0.90,0.83,0.79$, and
$0.73$, respectively.  To give a sense of the spread in the local
Fermi energies, we note that for the ideal trapped Fermi gas, the
ratio of the local Fermi energy at the average density to that at
the cloud center is approximately $0.6$.

\begin{acknowledgments}
We acknowledge funding from the NSF.  We thank the JILA BEC group
for discussions. DSJ acknowledges discussions with A. Kanigel at the Aspen Center for Physics.
\end{acknowledgments}


\bibliographystyle{prsty}


\end{document}